# Automatic alignment of an orbital angular momentum sorter in a transmission electron microscope using a convolution neural network

P. Rosi[1,2], A. Clausen[3], D. Weber[3], A.H. Tavabi[3], S. Frabboni[1,2], P. Tiemeijer[4], R.E. Dunin-Borkowski[3], E. Rotunno[1,*], V. Grillo[1]

1) Istituto Nanoscienze - CNR, via G. Campi 213/A, 41125 Modena, Italy
2) University of Modena and Reggio Emilia, via G. Campi 213/A, 41125 Modena, Italy
3) Ernst Ruska-Centre for Microscopy and Spectroscopy with Electrons, Forschungszentrum Jülich, 52425 Jülich, Germany
4) Thermo Fisher Scientific, PO Box 80066, 5600 KA Eindhoven, The Netherlands

*corresponding author address: *enzo.rotunno@nano.cnr.it*

**Abstract**

We report on the automatic alignment of a transmission electron microscope equipped with an orbital angular momentum sorter using a convolutional neural network. The neural network is able to control all relevant parameters of both the electron-optical setup of the microscope and the external voltage source of the sorter without input from the user. It is able to compensate for mechanical and optical misalignments of the sorter, in order to optimize its spectral resolution. The alignment is completed over a few frames and can be kept stable by making use of the fast fitting time of the neural network.

## 1-Introduction

Ever since the first electron microscope was built by Knoll and Ruska [1,2], transmission electron microscopes (TEMs) have evolved considerably. Even in their simpler realizations, they contain at least 10 lenses, 3 or more detectors and several apertures, which may be motorized. However, microscope alignment, which is required at the beginning of each experimental session, is still carried out primarily by the operator using the instrument control software. Furthermore, the advances that have resulted from the introduction of spherical aberration correctors [3–5], chromatic aberration correctors [6,7] and monochromators [8–11] which led to sub-atomic resolution imaging, have further increased the complexity of such instruments. The performance of the instrument thus entirely rely on the operators, making their skills the key factor during the experiments. Beside the instrument itself, also the experiments are becoming more and more complex and demand a high understanding and control of the system. An example of this challenge is illustrated by electron beam shaping, a rapidly advancing field in electron microscopy that aims to provide control over the electron beam wave function using amplitude or phase plates. These plates are normally fabricated using material-based holograms [12–16], but in recent years new phase plates based on microelectromechanical systems technology (MEMS) [17–20] have been developed. The latter devices can contain electrostatic elements such as metallic needles, toroids, or spirals whose electric field directly affects the electrons in the beam. By using a suitable distribution of electrostatic elements and stacking multiple phase plates, it is possible to impart almost any desired phase distribution to the electron beam [21]. An advantage of using MEMS technology is the ability to change in-operando the strength of each electrostatic element, in effect creating a programmable phase plate. Such concepts are often inspired by light optics, where their implementation is easier and more reliable, for example by using spatial light modulators. Recent developments in programmable structured light sources using molecular arrays [22] and plasmonic metasurfaces [23,24] are further promoting the concept of tailored illumination. MEMS technology should make such concepts more viable and accessible for electron microscopy, albeit at the cost of system complexity.

The complexity of electron columns and novel experimental setups is resulting in an ever higher demand for automation and instrument control, which is crucial to increase the speed, efficiency and reproducibility of electron optical alignment and to reduce the demand on the operator. Aberration correctors are currently controlled using semi-analytical models [25–28]. However, such models are not usually available for unconventional optics used in electron beam shaping. A more general approach is therefore required for microscope control. Recent advances in artificial intelligence (AI) in the automation field may be the answer for a general approach for experiments. This has prompted us to develop a convolution neural network (CNN) that is fit for purpose to aid us in diagnosing an orbital angular momentum (OAM) sorter for electron vortex beams [29], but in theory it can be generalized for a more complex system such as a full electron microscope.

An electron OAM sorter makes use of electron beam shaping to measure an electron beam's components of OAM in the propagation direction by decoupling the azimuthal and radial degrees of freedom of the electron beam. It also provides the first complete example of a lossless unitary base change that "diagonalizes" a quantum operator using wave manipulation. The implementation of an OAM sorter requires precise alignment of the electron microscope and control of two electron optical phase elements. However, no simple analytical model can be fitted to an OAM spectrum to measure the control parameters.

In this paper, we go beyond the proof of principle of our last paper and demonstrate the real time application directly on the microscope without user intervention. A neural network can be used to control a transmission electron microscope to achieve fast, reliable and stable alignment of the OAM spectrum of an electron beam. Since a large number of electron microscope parameters have to be tuned, we anticipate that a similar approach can be used in the future to automate other measurements in the TEM, as suggested by LeBeau and colleagues [30]. In light optics there have already been several examples in recent years [31–34], but even earlier [35], which showed how machine learning can be used for alignment/experimental tuning of operating parameters, further demonstrating that it is possible. Just as autonomous driving has been developed thanks to artificial intelligence, it should be possible to apply similar ideas to automatic alignment in electron microscopy.

**2- A Convolutional neural network for the electron Orbital Angular Momentum Sorter**

The CNN that we used to control the TEM has been designed specifically to evaluate the magnitudes of the primary parameters that need to be tuned for the alignment of an electrostatic OAM sorter [20,29]. Whereas details of the CNN are available in our previous paper [29], key elements such as network structure and type are described below.

As a reminder, an OAM sorter is a device that is able to analyse the quantum OAM components of a vortex beam [36]. In its most general realization, it comprises two phase-modifying elements. The first element is referred to as an unwrapper or "sorter 1" (S1), while the second element is referred to as a phase corrector or "sorter 2" (S2). The two phase elements must be in Fourier conjugate planes. The unwrapper performs a log-polar conformal transformation, which take the azimuthal phase gradient typical of a vortex beam and converts it into a linear phase gradient. It therefore performs a conformal coordinate transformation from $(x, y)$ to $(u, v)$, where $u$ and $v$ are coordinates in the Fourier conjugate plane of the unwrapper [36]. The phase shift introduced by the unwrapper is given by the expression

$$\varphi_{S1}(x,y) = \frac{ks}{f}\left(y \tan^{-1}\frac{y}{x} - x \log\left(\frac{\sqrt{x^2+y^2}}{L}\right) + x\right), \tag{1}$$

where $k = \frac{1}{\lambda}$ is the electron wave vector, $f$ is the focal distance between the two sorter elements and $s$ and $L$ are scaling parameters. The parameter $s$ determines the length of the transformed beam, while the parameter $L$ translates the transformed beam along $u$. For an electrostatic sorter, it is defined by the length of the charged (or biased) needle that is used as the unwrapper [36–38].

As the name suggests, the phase corrector corrects for phase inhomogeneities introduced by the unwrapper and prevents further S1-induced evolution of the electron beam upon propagation. Its phase shift is given by the expression

$$\varphi_{S2}(u,v) = -\frac{ksL}{f} \exp\left(-\frac{u}{s}\right) \cos\left(\frac{v}{s}\right). \tag{2}$$

We recently used MEMS technology to realize an *electrostatic* OAM sorter for electron vortex beams [20]. In this device, the unwrapper is an electrically biased needle, while the phase corrector takes the form of a series of alternately-biased electrodes. The electrostatic potential on each needle and electrode can be adjusted using an external voltage source. The value of $s$ in Eq. 1 and Eq. 2 can be changed by increasing the voltage applied to the needle in sorter 1, which increases the size of the beam in the S2 plane. The residual phase of the first element, after the transformation that it imparted to the electron beam, should then be matched perfectly by the phase of the second element. However, under the typical working conditions of the sorter, as reported in Ref. [20] and Ref. [29], the phase profiles of the elements vary rapidly in their respective planes. The alignment of the electron beam with respect to the second sorter element should therefore be nearly perfect, as a shift of a few nm can already reduce the resolution. The tedious and time-consuming precise alignment of the microscope and sorter elements required to obtain complete phase compensation and to optimize the sorting resolution in the final spectrum is what motivated us to develop and connect the CNN to the microscope, which should allow for a fast, precise and reliable alignment. As described in Ref. [29], the main misalignments are rotation of the unwrapped beam with respect to the phase corrector, size mismatch between the unwrapped beam and the periodicity of sorter 2, defocus of the beam prior to being transformed and incorrect positioning of the unwrapped beam in the phase corrector plane.

The CNN that was connected to the microscope and voltage generator is described in Ref. [29]. It is composed of 13 layers, as shown schematically in Figure 1: an input layer (a 128 x 128 pixel image), 5 convolution layer filters (each of which is followed by an average-pooling layer filter) and two fully connected layers leading to the output, corresponding to a total of 2,480,326 trainable parameters. The chosen activation function was the rectified linear unit (ReLU), the learning algorithm used was Adam [39], the learning rate was 0.001, while the loss function was the root mean square difference between the predicted and true misalignment coefficients. The CNN was implemented using the Keras library [40] and the TensorFlow backend [41]. It was trained on 20 000 simulated images (+ 2000 images for validation) for random values of 6 misalignment parameters. Details of how the simulated images were obtained are reported in Ref. [29]. It should be noted that the CNN did not converge to a correct value without accounting for decoherence. The decoherence effect was accounted for by fitting a classical Gaussian model directly to experimental images [42,43], with different values of the coherence length used to find the correct one. The best-performing model had a gaussian width of $0.45\hbar$ in the OAM spectrum.

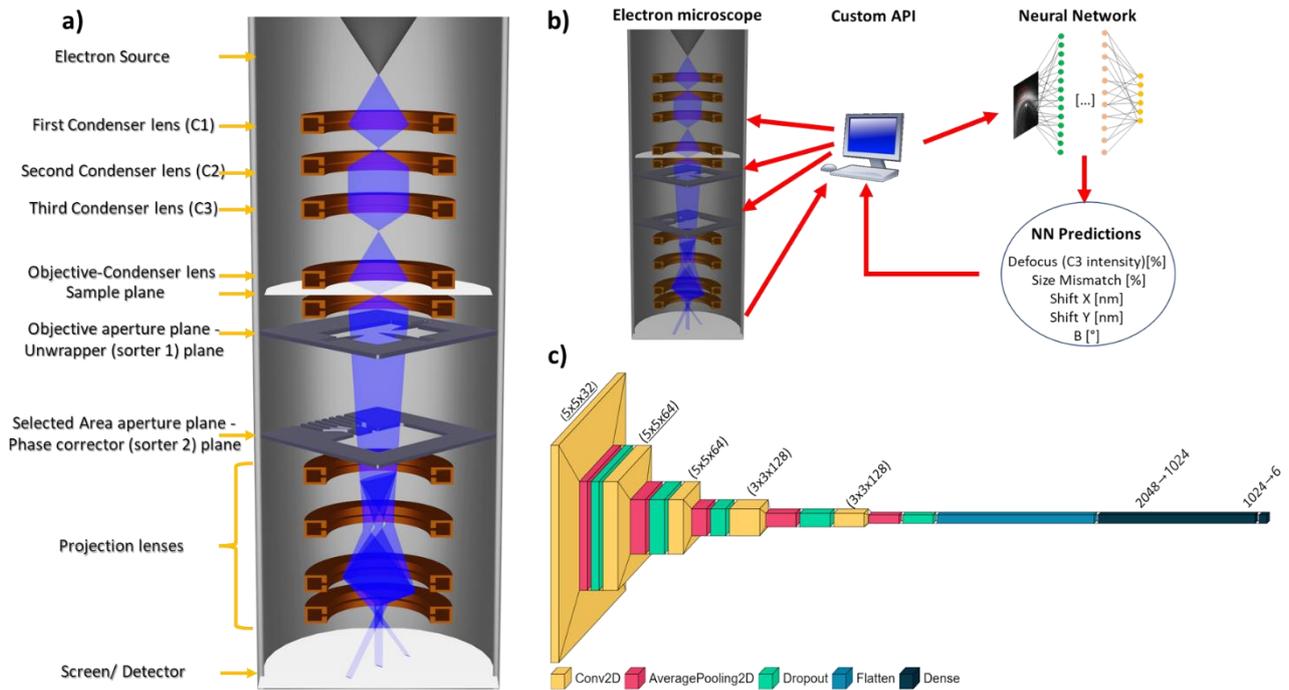

*Figure 1: Here are shown a schematic view of the Transmission Electron Microscope electron-optical configuration (a), a scheme showing how our control of the TEM through the CNN is achieved (b), while lastly the layout of the convolution neural network is reported (c). Further details about the network structure can be found in Ref.* [29].

## 3-Control loop

In order to achieve real-time automatic alignment, image data from the microscope was fed into the CNN, then the output from the CNN was converted to control commands, which were fed to the microscope and connected devices *via* application programming interfaces (APIs). The CNN and control loop run on a custom python3 script. As an overview, the control loop can be described by the following steps:

1. Acquisition of an image using the K2 camera.

2. Pre-processing of the image.

3. Feeding of the pre-processed image into the pre-trained network.

4. Determination by the network of the parameters that it expects to perfectly align the sorter.

5. Calculation by the software of changes to the alignment parameters required for optimal alignment.

6. Application of the corrections as changes to the microscope parameters and to the voltage sources.

*3.1-Image data acquisition*

Experiments were performed on an FEI Titan G2 60-300 TEM [44] in Forschungszentrum Jülich. For data acquisition, a K2 camera was operated in *In Situ* (IS) mode. A prototype of a K2-IS adapter for the LiberTEM-live framework [45] was used to perform continuous data capture and analysis at the full data rate of the K2-IS camera (400 fps at 2048 x 1860 pixels). In the present work, sets of 400 frames were summed to create 1 second exposure images at each control step. In order to work with LiberTEM-live, a processing server running the Debian GNU/Linux "Buster" distribution was connected to the data switch of the K2 camera using two 10 GBit network interfaces. The K2 data switch was reconfigured to forward all multicast UDP datagrams coming from the K2 digitizers to the processing server. The prototype K2 IS adapter receives, decodes and sorts the UDP datagrams and executes arbitrary processing and analysis functions (referred to as UDFs in LiberTEM) on this data stream, allowing for fast visual feedback and near-real-time feedback loops. The LiberTEM back-end and LiberTEM-live separate technical aspects, such as communication with the camera, synchronization and decoding/re-ordering of the data streams, from data processing in the UDF. In this way, a UDF can run unmodified on any data stream from a LiberTEM-compatible detector or file format.

*3.2-Pre-processing*

The recorded images were pre-processed to match images in the training dataset in size and orientation.

Precise calibration is crucial for the CNN to provide quantitative results. For this purpose, calibration spectra were obtained using petal beams, which have known OAM decompositions and can be produced easily using synthetic holograms made by focused ion beam milling. A review on this subject and on beam shaping using synthetic holograms is available in Ref. [46].

Once the pixel size of the experimental images had been measured, they could be scaled, cropped and normalized over the interval [0,1] to match the simulations in the training dataset.

As the model is not rotationally invariant, the rotation of the spectrum also had to be calibrated. Whereas the OAM axis is horizontal in the simulations, in the experimental images it is defined by the orientation of the second element of the sorter with respect to the camera and by the excitation of the lenses in between them. As the OAM spectrum is dispersed along a single direction, calibration of the rotation is straightforward.

*3.3-Inference*

The network calculates the corrections that it expects to provide a perfect alignment of the sorter. These settings are applied relative to the current state of the system. A damping factor and a rate limit are used to take calibration errors into account and to manage interdependence between the parameters. As demonstrated in our previous work [29], the network output corrections have a linear trend. Damping prevents overshooting the target by limiting, in most of the cases, the jumps around the convergence point. Such jumps are expected since the network was trained with limited resolution. A damping factor of 0.4 was used for the beam shift (X/Y) and intensity. A smaller damping factor of 0.2 was used for the scaling factor *s* of sorter S1, being it the most sensitive parameter.

Custom calibrations were performed to provide a conversion from the damped output of the network to the units used by the microscope API. The beam shift (X/Y) had to be flipped in the X and Y directions (defined in the sorter 2 plane, where the positive direction of X moves towards S2 and Y is defined such that the right-hand rule applies), rotated (because the Cartesian coordinate system of the network is rotated with respect to the Cartesian coordinate system of the microscope) and scaled from the network output to the microscope units. The excitation values of condenser lens C3 also had to be scaled and an offset added to them. This calibration was determined by adjusting the microscope in such a way that the effect of the control inputs could be observed relative to the sorter device.

*3.4-Instrument control*

Multiple devices were controlled during the experiment. A 16-channel voltage source (Stahl-Electronics' DC precision voltage source BS-series) was used to power the two sorter devices. The TEM parameters (beam shift (X/Y), C3 intensity and other parameters) were controlled *via* the COM scripting interface of the microscope. All of the devices, with the exception of the K2 camera, were controlled using a Tango [47] setup, which allows network-transparent control of different devices and encapsulates the actual control details in the tango device server. After the device server was set up, it could be used by installing a Tango client library from any of the supported languages (currently: LabView, Matlab/Octave, Java, Python, C++ among others) or *via* a REST API. On the client side, there are no further hardware-specific dependencies.

## 4-Automatic control

We performed a series of test runs using the described setup. In Figure 2 is reported a series of images where on purpose we started from incorrect condition and the CNN was then asked to correct the OAM spectrum. As already underlined in paragraph 3.1 the camera is used to perform a continuous data capture to create 1 second exposure images. Only three of the main alignment parameters that act on the OAM sorter were controlled: size mismatch (*i.e.*, voltage applied to the main needle of the unwrapper) and beam shift (X/Y).

Estimated values of the parameters were obtained by re-analysing the image series with the same CNN that was used to control the microscope. Figure 3 shows the time evolution of the values corrected by the CNN. Each of the five series shown in Fig. 3 involved starting from a random condition (by using software to randomly change the correctable parameters within the correction range after manually correcting the OAM spectrum), except for the starting value of the bias applied to sorter 1 (which was increased from one series to the next to see if the starting condition of sorter 1 had an effect on the final value). In each series, the three parameters that were corrected reached values of zero (meaning that it was not necessary to compensate for them further). This situation was achieved, on average, after 7 frames. The values then oscillated about zero, while maintaining stable spectra. For most of the series, the final stable value of the sorter 1 bias was 5.9 V, while for a couple of them it was 6.05 V, which is within the 2% tolerance that was defined in the previous paper (and therefore the degree of accuracy of the network).

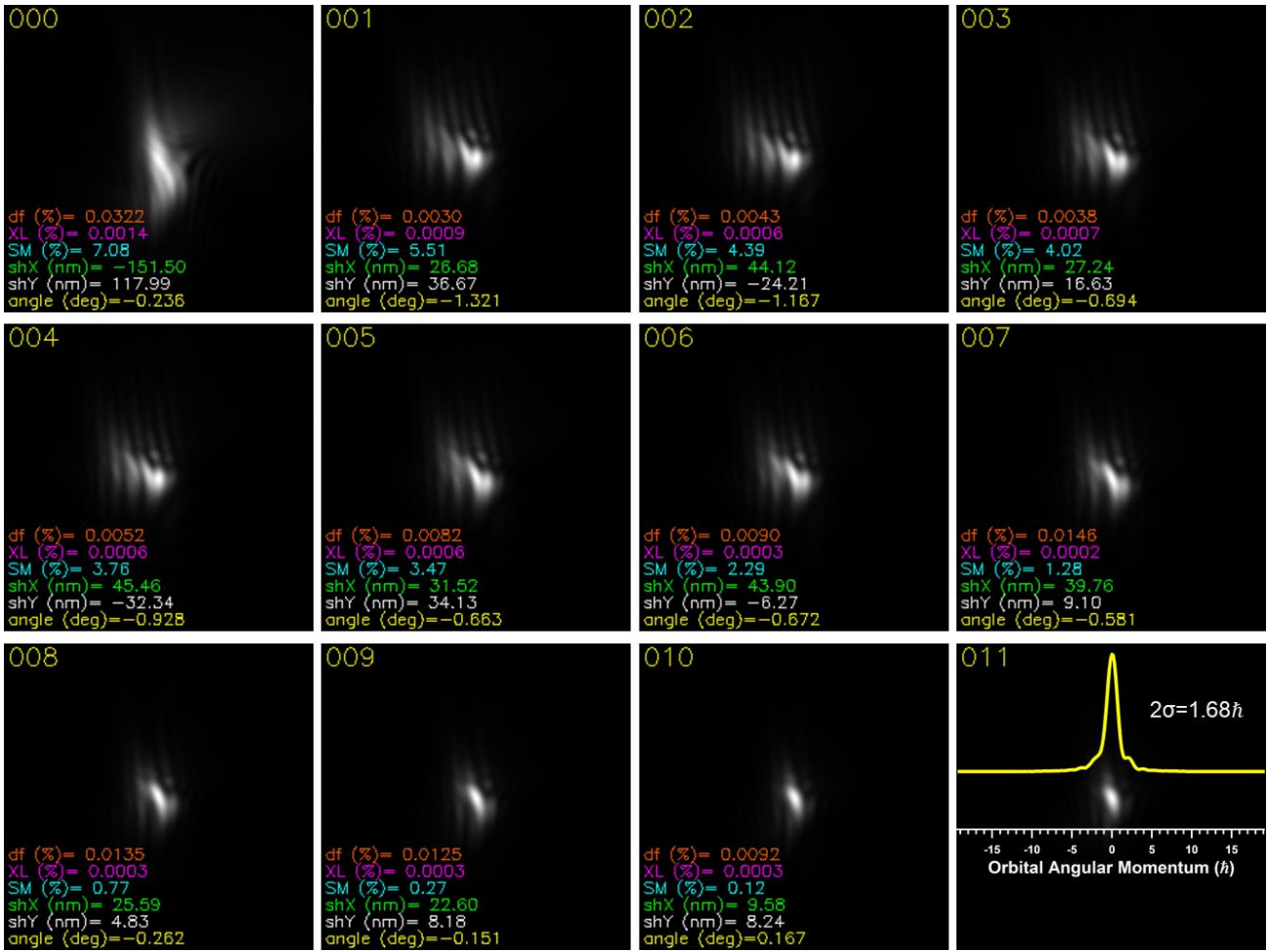

Figure 2: Experimental image series showing the evolution of an OAM spectrum in time as the CNN provides correction values. Each image is a 1 second exposure image and init, the upper left corner shows the frame number (i.e., assessing a time stamp in seconds from the frame 0 ), the lower left corner shows the value of the alignment parameters estimated by the CNN, which were fed to the microscope to generate the images shown in the subsequent frames. The final frame shows a line scan, in which the width of the gaussian profile, which corresponds to twice the standard deviation, takes a value of 1.68ℏ.

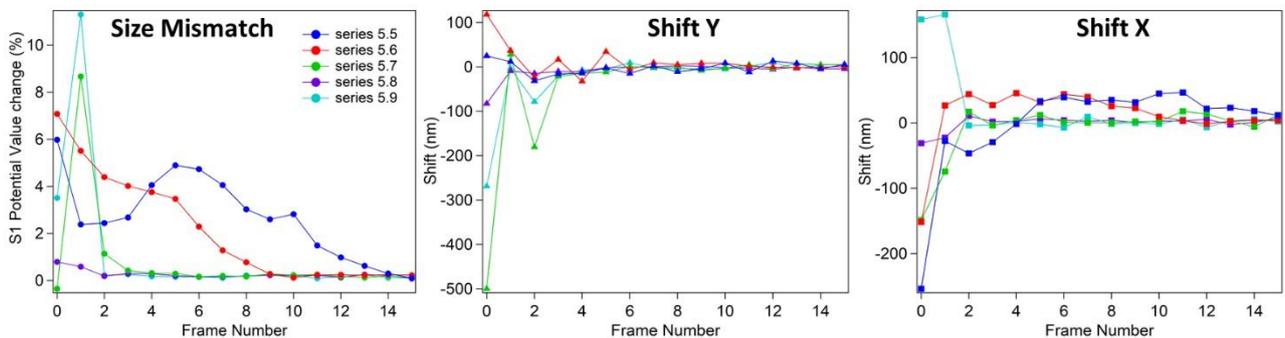

Figure 3: Behaviour of CNN output as a function of frame number showing size mismatch and beam shift (X/Y). Each parameter tends to zero after a few iterations.

We then tried to control a fourth parameter: the sorting optics defocus, which is controlled by the excitation of the C3 lens of the microscope. Figure 4 shows an image series that illustrates how the OAM spectrum is corrected and improved over time.

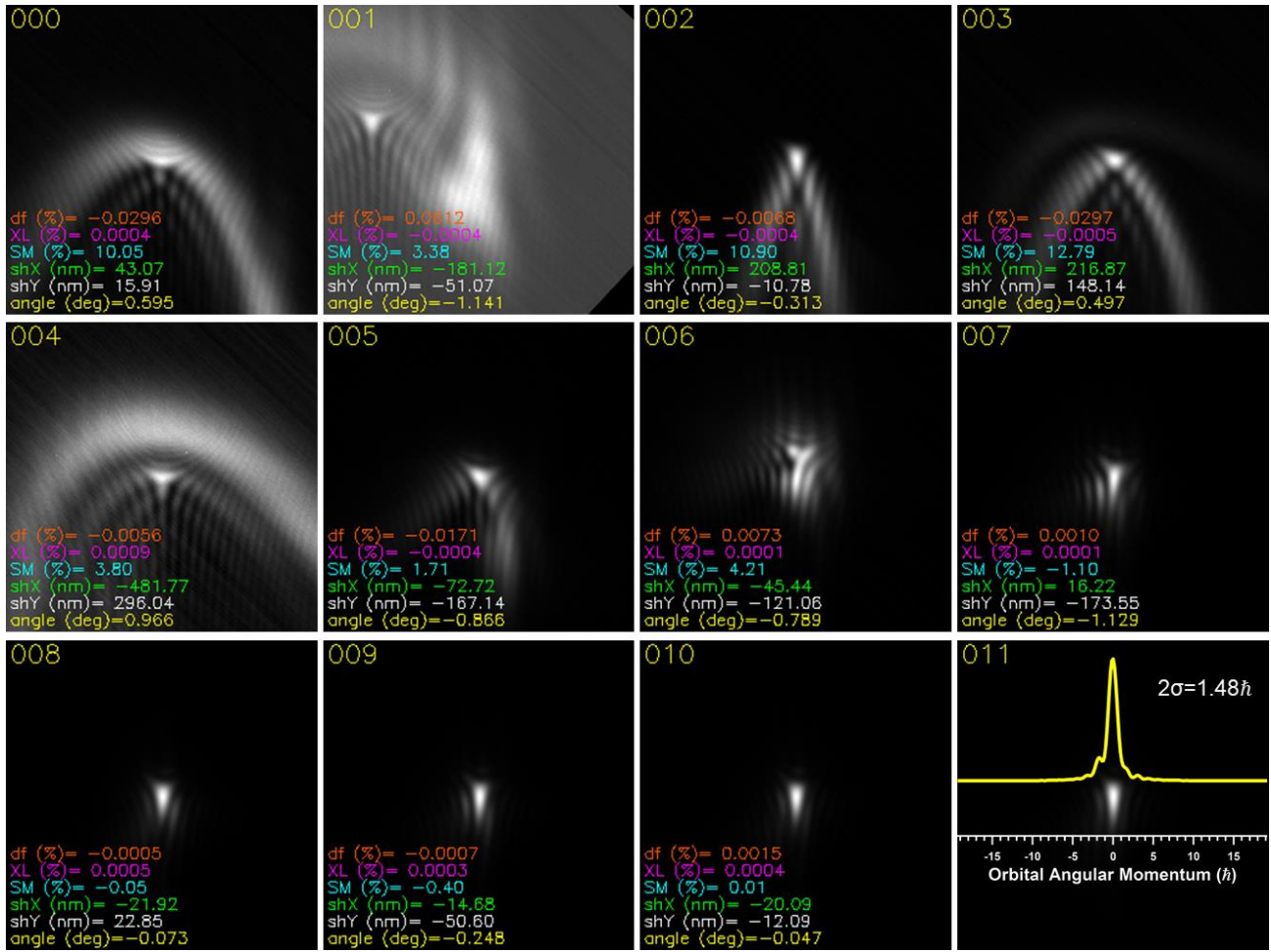

Figure 4: Experimental image series showing how an OAM spectrum evolves in time as the CNN provides correction values for size mismatch, beam shift (X/Y) and defocus. An initial "wrong" OAM spectrum was corrected by the CNN. The final frame shows a line scan, in which the width of the gaussian profile, which corresponds to twice the standard deviation, takes a value of 1.48ℏ.

It is interesting to observe how robust the CNN is in Fig. 4. In the second frame, either there was a problem during image acquisition due to the beam being near to the border of the K2 sensor or the spectrum was corrected incorrectly. However, from the third frame the CNN is able to recover (even though it was fed with only partially correct data from the previous frame) and gradually compensates for the misalignments until it reaches a stable condition. This behaviour is confirmed by observing trends of the estimated values provided by the CNN, as shown in Figure 5. The trends were obtained by re-analysing the image series using the same CNN that was used to control the microscope.

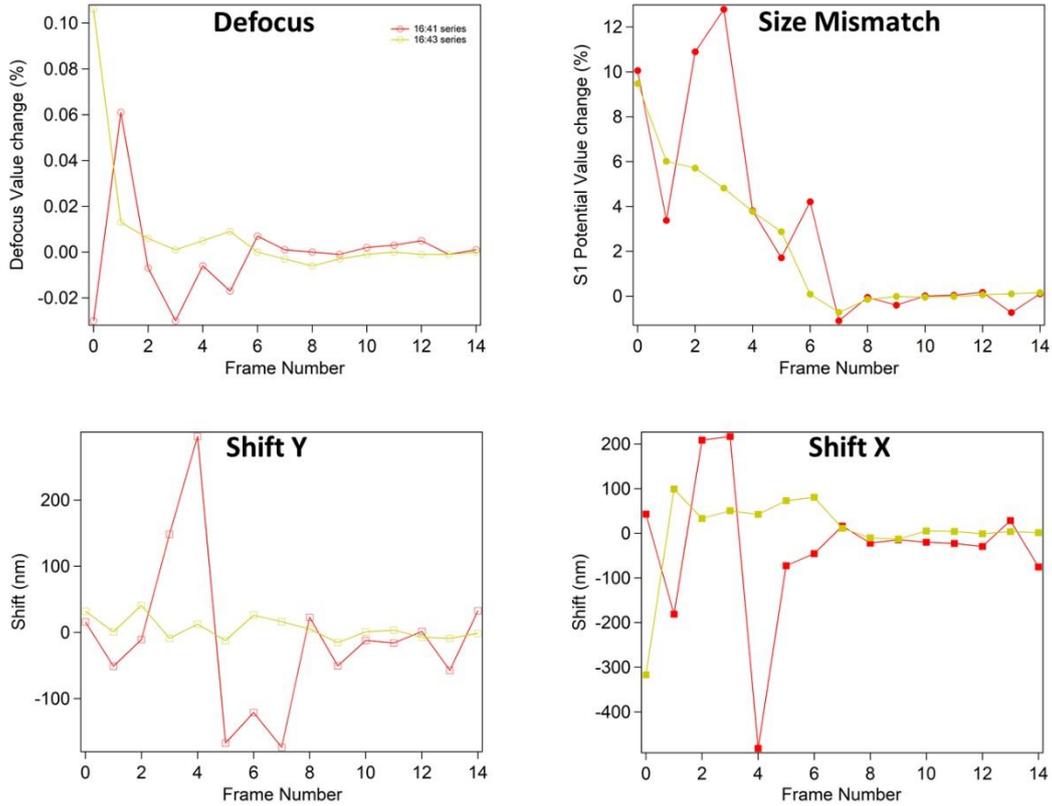

*Figure 5: Behaviour of CNN output as a function of frame number showing size mismatch, beam shift (X/Y) and defocus. Each parameter tends to zero.*

Although the series shown in Fig. 3 were obtained starting from a sorter 1 potential that was chosen by us and incremented at the start of each series to assess whether it had an effect on the results, while a random sorter 1 potential was used for the two series shown in Fig. 5.

The parameter that is commonly used to benchmark an OAM sorter is its sorting resolution. By making use of a prior pixel calibration of the OAM spectrum image, the measured sorting resolution in the final frame of Fig. 4 is determined to be approximately $1.5\hbar$, which is comparable to that obtained in Ref. [20]. This is a remarkable result, since the software-guided alignment takes only a few seconds to reach a stable condition and is then able to keep this condition stable, thereby overcoming possible device or microscope instabilities. In contrast, an operator usually requires minutes to reach a comparable or better result. For further improvement in the sorting resolution, control over more TEM parameters is necessary. Moreover, as previously mentioned, the CNN was trained on simulated images that are based upon an ideal mathematical model of the sorter, resulting in limited accuracy of the network when applied to a real situation. Refinement of the network by training it directly at the microscope using reinforcement learning schemes is expected to reduce the number of frames that are required to tune the device. Improvements in the tuning speed will allow us to perform a live correction of the misalignments where the user will hardly notice it happening.

**Conclusions**

In this paper, a convolutional neural network has been used to directly ( i.e., without intervention of the operator) control the hardware part of the optical system of a transmission electron microscope and an external generator that controls a programmable electrostatic phase plate operated as an orbital angular momentum sorter. The neural network is robust enough to correct for misalignments that affect the experimental OAM spectrum, even though it was trained on simulated spectra. It can also be used to overcome unexpected events, such as a sudden change in defocus. Tuning of the microscope optics and external power supply takes on average less than 10 s for full correction of a spectrum, providing a final orbital angular momentum resolution of approximately $1.5\hbar$. This is a major improvement over manual correction by the microscope operator, which typically takes several minutes and requires prior experience. In principle, this approach can be extended to other devices in the column, including correctors and monochromators, in the future.

**Acknowledgments**

This project received funding from the European Union's Horizon 2020 research and innovation programme (Grant No. 766970, project "Q-SORT"; Grant No. 856538, project "3D MAGiC"; Grant No. 823717, project "ESTEEM3"; Grant No. 101035013, project "MINEON"), from the Deutsche Forschungsgemeinschaft (Project-ID 405553726 – TRR 270) and from the DARPA TEE program (Grant MIPR# HR0011831554).

**Data availability statement**

The data and custom API that support the findings of this study are openly available in Zenodo at http://doi.org/10.5281/zenodo.5715058 and 10.5281/zenodo.6420912, respectively.

**Supplementary material**

*Settings for live processing of K2 IS data*

In order to ensure proper communication and reliable data delivery, several network and Linux kernel parameters must be properly tuned:

- As the UDP datagrams are encapsulated into ethernet jumbo frames, an appropriate MTU needs to be set, for example 9000. Failure to do so will result in an increase in *rx_length_errors* statistics in the output of *ethtool -S <ifname>*;
- The UDP datagrams arrive as multiple IP fragments, which need to be assembled into whole packets by the operating system. Without adjusting the system defaults, this process may require too much memory and result in dropped frames. The available buffer memory can be changed by writing to */proc/sys/net/ipv4/ipfrag_high_thresh*. A suitable process involves the observation of assembly failures under load with *netstat -s | grep -i assem* and increasing the fragmentation threshold until they no longer appear.
- Under high load, it is possible that the UDP receives a buffer overflow, which is another possible cause of dropped packets and can be observed via netstat -s, where the "receive buffer errors" counter will increase. For the individual sockets, it can be observed by tracking the d column in *ss -unlp -m -A udp*. In order to decrease the likelihood of dropped packets, it is possible to resize the buffers by increasing */proc/sys/net/core/rmem_default* and */proc/sys/net/core/rmem_max*.
- Depending on the network hardware, it can be beneficial to set up RX (receive) queue hashing, in order to better distribute the network load across multiple CPU cores. For example, on supported network hardware, by using *ethtool -N <ifname> rx-flow-hash udp4 sdfn* the received network traffic is routed to different RX queues and therefore different CPU cores, depending on the source IP address, the destination IP address, the source port and the destination port [48].